\documentclass[10pt,aps,prd,twocolumn,floats,floatfix,showpacs,superscriptaddress,nofootinbib]{revtex4-1}

\makeatletter
\def\l@subsubsection#1#2{}
\def\l@subsubsubsection#1#2{}
\makeatother

\usepackage{amsfonts}
\usepackage{graphicx,amssymb,amsmath,amsthm,epsfig,epsf}
\usepackage[usenames]{color}
\usepackage{epstopdf}

\usepackage{aas_macros}
\usepackage{bm}
\usepackage{dcolumn}
\usepackage[utf8]{inputenc}
\usepackage{latexsym}
\usepackage{rotating}
\usepackage{longtable}

\usepackage{enumerate}
\usepackage{tensor,multirow}
\usepackage{url}
\usepackage[linktocpage]{hyperref}
\usepackage{diagbox}
\usepackage{colortbl}
\usepackage{tikz}
\usetikzlibrary{patterns}
\usepackage{tabularx}

\def\nn{\nonumber}

\def\be{\begin{equation}}
\def\ee{\end{equation}}
\newcommand{\beq}{\begin{eqnarray}}
\newcommand{\eeq}{\end{eqnarray}}

\def\ba{\begin{align}}
\def\ea{\end{align}}

\newcolumntype{Y}{>{\centering\arraybackslash}X}

\newlength{\flexcheckerboardsize}

	\newcommand{\defineflexcheckerboard}[5]{
    \setlength{\flexcheckerboardsize}{#2}
    \pgfdeclarepatterninherentlycolored{#1}
        {\pgfpointorigin}{\pgfqpoint{2\flexcheckerboardsize}    
        {2\flexcheckerboardsize}}
        {\pgfqpoint{2\flexcheckerboardsize}
        {2\flexcheckerboardsize}}%
        {
            \pgfsetfillcolor{#4}
            \pgfpathrectangle{\pgfpointorigin}{
            \pgfqpoint{2.1\flexcheckerboardsize}    
                {2.1\flexcheckerboardsize}}
          \pgfusepath{fill}
          \pgfsetfillcolor{#3}
          \pgfpathrectangle{\pgfpointorigin}
            {\pgfqpoint{\flexcheckerboardsize}
            {\flexcheckerboardsize}}
          \pgfpathrectangle{\pgfqpoint{\flexcheckerboardsize}
            {\flexcheckerboardsize}}
            {\pgfqpoint{\flexcheckerboardsize}
            {\flexcheckerboardsize}}
            \pgfusepath{fill}
        }
}

\defineflexcheckerboard{flexcheckerboard_bw}{.5mm}{black}{white}{0}
\defineflexcheckerboard{flexcheckerboard_redblue}{2mm}{red}{blue}{0.1}
\defineflexcheckerboard{flexcheckerboard_greenorange}{10mm}{green}{orange}{0}
\defineflexcheckerboard{flexcheckerboard_bluecyan}{.2mm}{cyan}{blue}{0}

\begin{document}
\title{Axisymmetric deformations of neutron stars and gravitational-wave astronomy}

\author{Guilherme Raposo}
	\email{guilherme.raposo@roma1.infn.it}
	\affiliation{Dipartimento di Fisica, ``Sapienza" Universit\`a di Roma \& Sezione INFN Roma1, Piazzale Aldo Moro 
5, 00185, Roma, Italy}

\author{Paolo Pani}
	\email{paolo.pani@uniroma1.it}
	\affiliation{Dipartimento di Fisica, ``Sapienza" Universit\`a di Roma \& Sezione INFN Roma1, Piazzale Aldo Moro 
5, 00185, Roma, Italy}

\begin{abstract}
Einstein's theory of general relativity predicts that the only stationary configuration 
of an isolated black hole is the Kerr spacetime, which has a unique multipolar structure 
and a spherical shape when nonspinning. This is in 
striking contrast to the case of other self-gravitating objects,
which instead can in principle have arbitrary deformations even in the static case. 
Here we develop a general perturbative framework to construct stationary stars with 
small axisymmetric deformations, and study explicitly compact stars with an intrinsic 
quadrupole moment. The latter can be sustained, for instance, by crust stresses or strong magnetic fields.
While our framework is general, we focus on quadrupolar deformations of neutron stars induced by an anisotropic crust, 
which continuously connect to spherical neutron stars in the isotropic limit.
Deformed neutron stars might provide a more accurate description 
for isolated and binary compact objects, and can be used to improve constraints on the neutron-star equation of state 
through gravitational-wave detections and through the observation of low-mass X-ray binaries.
We argue that, for values of the dimensionless intrinsic quadrupole moment of few 
percent or higher (which can be sustained by an elastic crust with ordinary parameters), 
the effect of the deformation is stronger than that of tidal interactions in coalescing neutron-star binaries,
and might also significantly affect the electromagnetic signal from accreting neutron stars.
Current observational bounds on the post-Newtonian coefficients in the gravitational waveform signal from GW170817 do 
not exclude that the neutron stars in the binary had some significant intrinsic deformation.
\end{abstract}

\maketitle


\section{Introduction}
Neutron stars~(NSs) harbor the highest densities, the strongest magnetic fields, the 
highest binding energy per nucleon, and the strongest spacetime curvatures in the 
universe. Provided their interior and spacetime can be accurately modeled using nuclear 
physics and general relativity, NSs are unique probes of all fundamental interactions and 
ideal laboratories to test foundational physics and high-energy 
astrophysics~\cite{Lattimer:2004pg}.

However, at variance with black holes, NSs are not simple objects.
Within Einstein's theory of general relativity, the black-hole uniqueness 
theorems imply that the ultimate stationary outcome of the 
gravitational collapse must be a Kerr black hole~\cite{Carter71,Hawking:1973uf}. The 
latter has an infinite number of multipole moments~\cite{Geroch:1970cd} which are anyway 
uniquely determined in terms of its mass and angular momentum~\cite{Hansen:1974zz}. 
When nonspinning, any isolated black hole in the universe must be spherically symmetric and 
described by the Schwarzschild spacetime.

This remarkable simplicity does not hold true for other self-gravitating 
objects, in particular for NSs. There is no compelling reason preventing NSs 
to be arbitrarily deformed away from spherical symmetry, even when nonspinning.
In fact, one might even argue the opposite, namely that spherical symmetry is a mere 
idealization and that all astrophysical formation processes are intrinsically 
\emph{asymmetric}, e.g., due to magnetic fields, environmental effects, crust shears, elasticity, turbulence,
convective instabilities, collimated neutrino 
fluxes, gravitational-wave~(GW) emission, kicks, deformations of the progenitor proto-NS, 
etc. It is therefore natural to expect that a newly-born NS might be deformed to some degree and 
that it can reach a stationary, \emph{axisymmetric} (but not necessarily spherical) configuration through GW 
emission~\cite{ShapiroTeukolsky} over long time scales.

In this paper we are not interested to model the complex physical mechanisms (e.g., core-collapse 
supernovae~\cite{1966ApJ...143..626C,Janka:2006fh} or a compact-binary 
coalescence~\cite{Shibata:2005ss,GW170817,2008PhRvD..78h4033B}) that may lead to the formation of a highly deformed 
stars, but rather we wish to understand whether realistic matter can sustain \emph{stationary} stars which 
deviate from spherical symmetry. While it is possible that the processes mention above lead to large deviations from 
sphericity, we shall work in a perturbative regime and treat these deviations as small perturbations around a spherical 
star. This approximation is also consistent with the fact that --~to the best of our knowledge~-- these deformations 
have been so far neglected when modelling stationary sources, e.g. old NS in low-mass X-ray binaries or in coalescing 
binaries.

Non-axisymmetric deformations have been intensively studied as a source of 
quasi-monochromatic GWs from isolated spinning 
NSs~\cite{Andersson:1998ze,Glampedakis:2017nqy,2012PhRvL.109h1103G}. Strong constraints exist on departures from 
axisymmetry, typically measured by the ellipticity of a 
NS~\cite{Andersson:1998ze,Ushomirsky:2000ax,Haskell:2006sv,Abbott:2018qee}.
Additionally, axisymmetric deformations have also been studied in the past, especially in the 
context of magnetized~\cite{Ioka:2003nh,Colaiuda:2007br,2012MNRAS.427.3406F,2014MNRAS.439.3541P,2015MNRAS.447.3278B, 
1995A&A...301..757B} or spinning~\cite{Hartle:1967he,HT} stars.
However, in the absence of magnetic fields or rotation, deformations can only be supported by the elastic properties of 
the material. The elastic properties of NSs are under active scrutiny, both in the stellar
core~\cite{Rajagopal_2006,Rajagopal_2006b,Alford_2008} and in the crust (see, e.g., Ref.~\cite{Chamel:2008ca} for a 
review on the topic). Although axisymmetric deformations do not destabilize the star through 
GW emission\footnote{GW emission from axisymmetric NSs can occur only if 
the stellar angular momentum is misaligned with the axis of symmetry, leading to 
precession. We focus here on stationary configurations, for which the angular momentum (if present)
coincides with the axis of symmetry.}, they might give rise to important effects 
in isolated and binary NSs. The scope of this work is to discuss this scenario.
Unless otherwise stated, we use $G=c=1$ units henceforth.

\begin{figure*}[ht]
\centering
\includegraphics[width=0.6\textwidth]{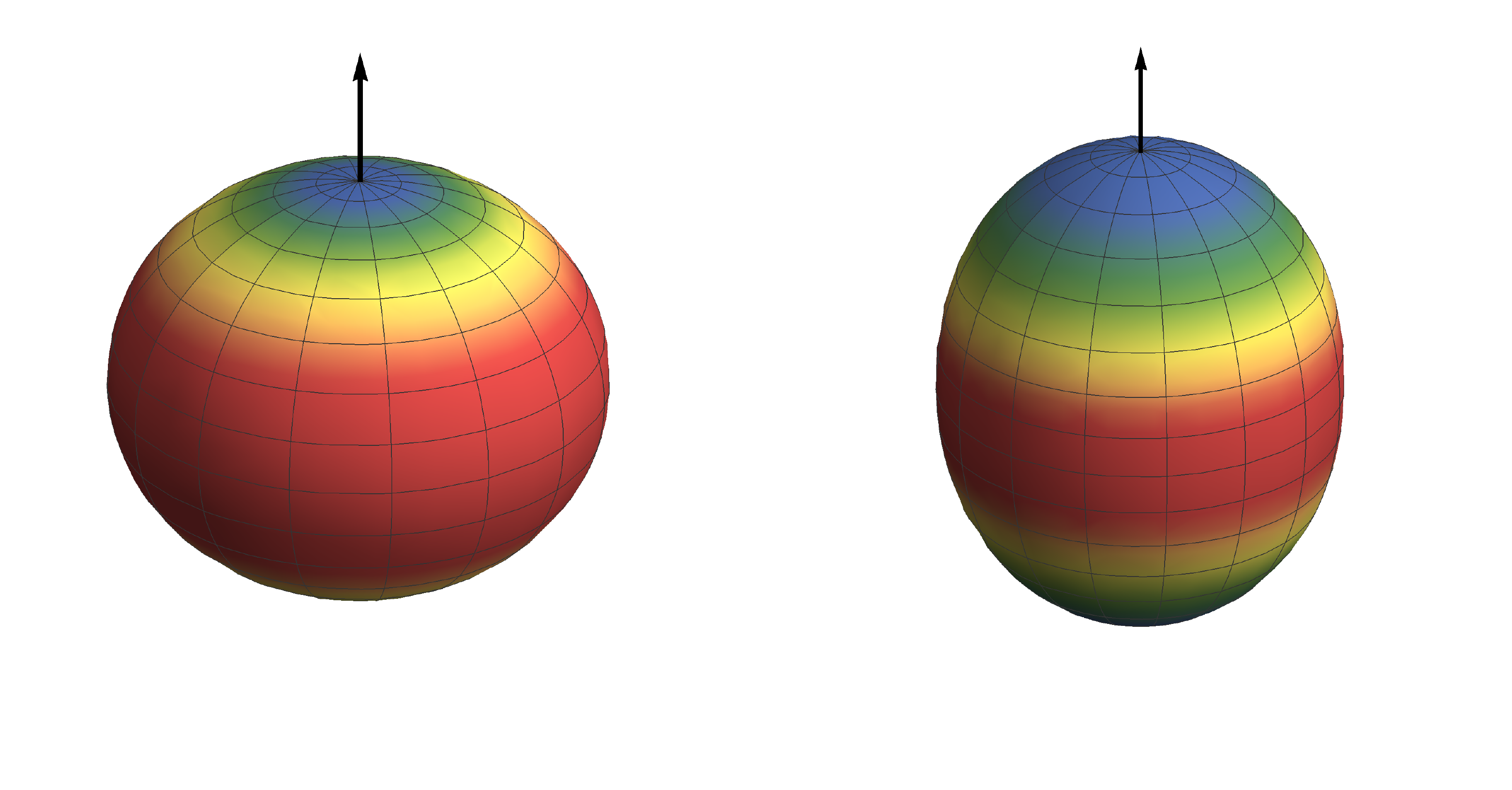}
\caption{{\bf Deformed NSs.} Illustrative embedding diagrams of a deformed NS. The arrow 
is the angular momentum vector and (if present) coincides with the axis of symmetry. The colors are 
weighted to represent the current multipole moment. In this example the body on the left 
is oblate due to a negative intrinsic quadrupole moment which adds to the the spin-induced 
term (the latter contributes to make the star oblate), whereas the body on the right 
is prolate because the effect of a positive intrinsic quadrupole moment is stronger 
than the spin contribution.
}
\label{fig:embedding}
\end{figure*}

\begin{figure*}[th]
\centering
\includegraphics[width=0.475\textwidth]{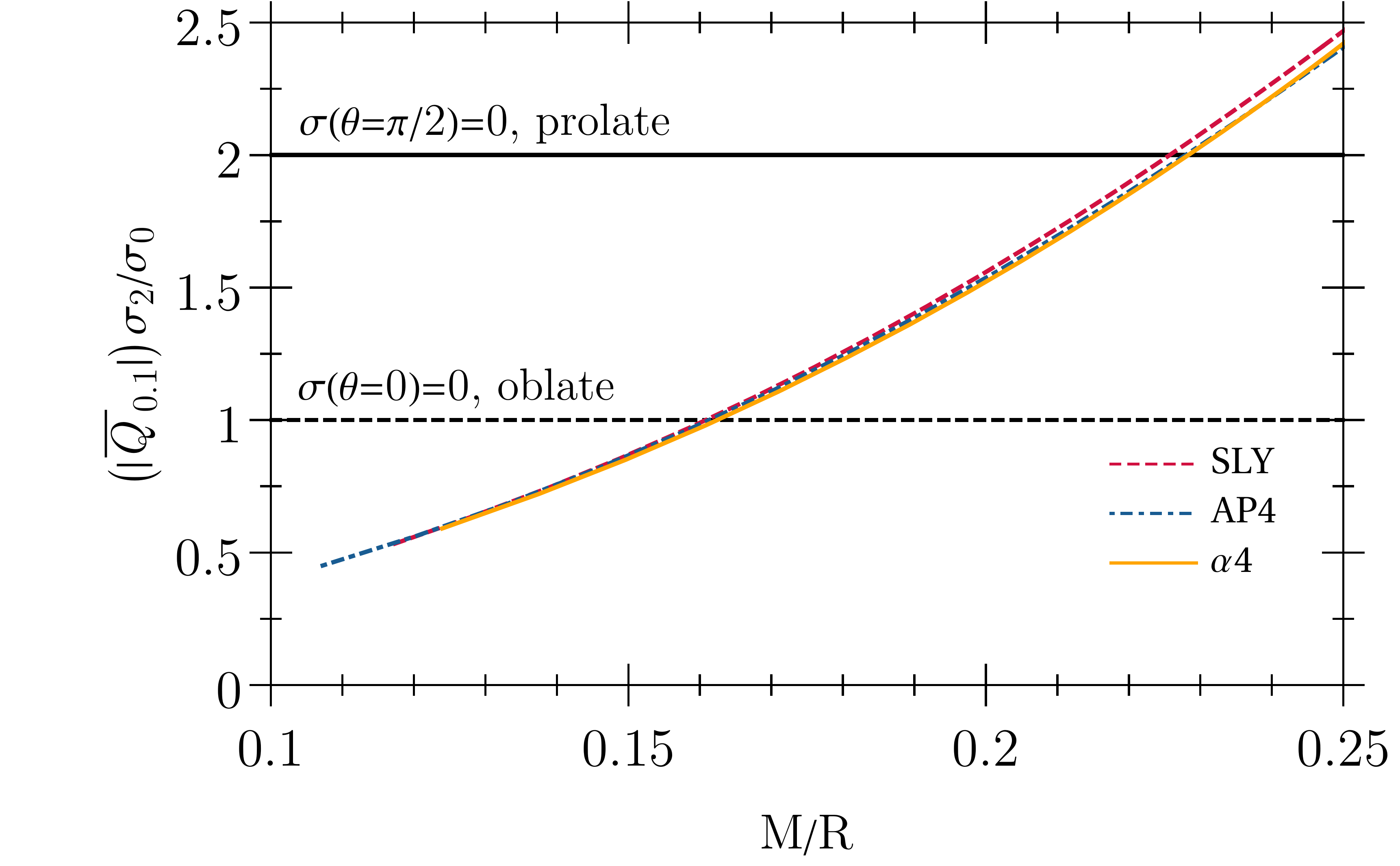}
\includegraphics[width=0.475\textwidth]{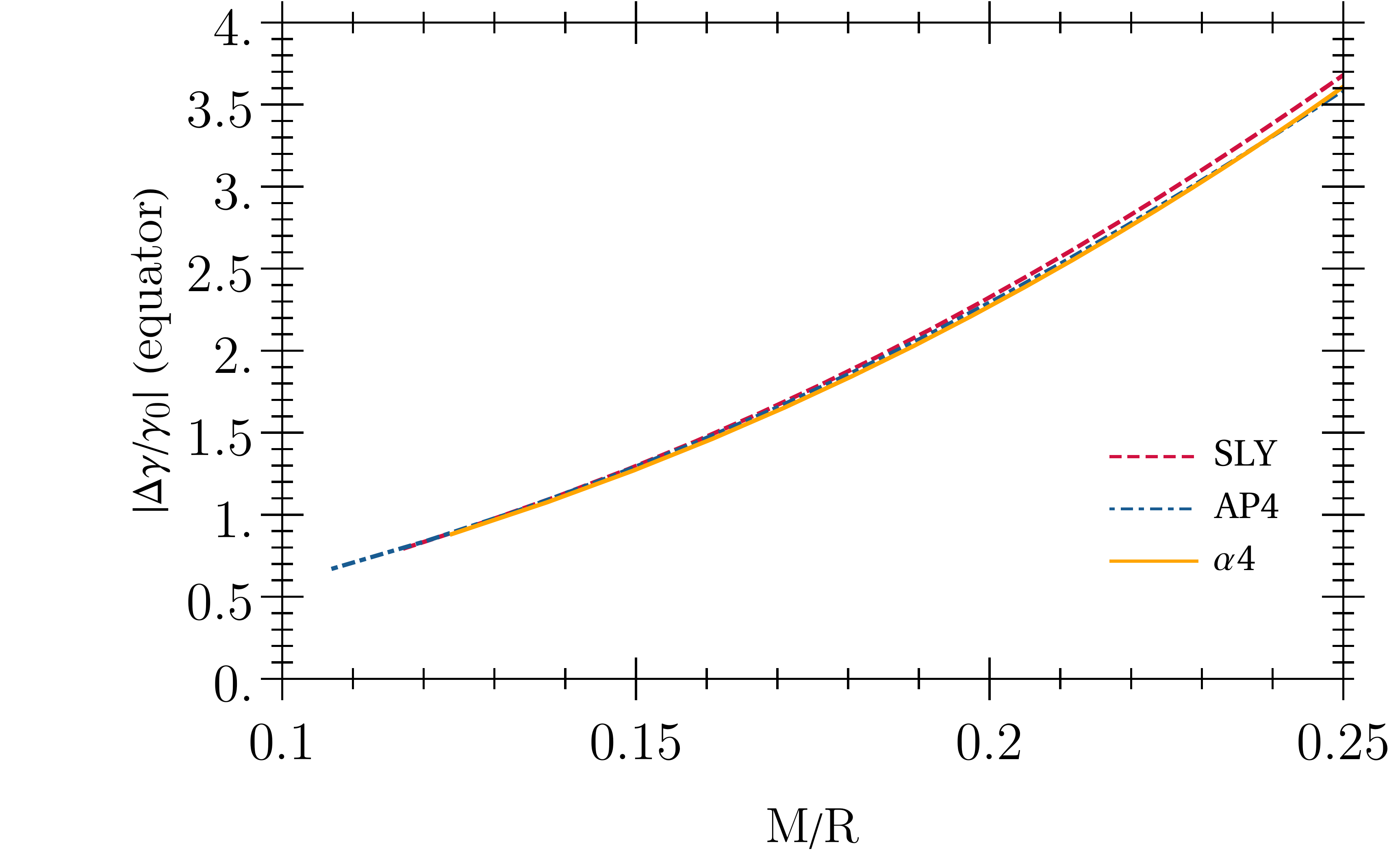}
\caption{{\bf Properties of the crust thin-shell.}  \emph{Left panel:} quadrupolar surface density $\bar{Q}\sigma_2$ 
obtained with different tabulated EoS. This quantity is normalized by a quadrupole value $|\bar{Q}|=0.1$ and by a 
typical reference value for the spherical surface density of the shell $\sigma_0$, corresponding to $1\%$ of the NS 
mass. Due to the quadrupolar deformation of the star, the density can be negative depending on the ratio between the 
magnitude of the quadrupolar density deformation and the spherical shell density. For prolate (oblate) configurations, 
the minimum value of the density occurs at the equator (poles) and the limit of zero density is represented by the  
horizontal solid (dashed) black line. When the ratio is below the black line, the surface density is positive 
everywhere, whereas it can be negative for values above the curve. 
\emph{Right panel:} surface anisotropy $\Delta \gamma \equiv \gamma_\theta-\gamma_\phi$ at the equator, normalized by 
${\bar Q}=0.1$ and by the pressure of the spherical shell, $\gamma_0$, corresponding to a crust with $1\%$ of the NS 
mass. The quadrupolar deformation of the star induces anisotropy in the surface pressure of the star between the two 
angular directions. The relative magnitude is compatible with the maximum one allowed by elastic 
shears~\cite{Baiko:2018jax}. 
Note that both $\sigma_2$ and $\Delta\gamma$ grow monotonically with the compactness of the 
star and are almost independent of the EoS.
}
\label{fig:sigma2}
\end{figure*}

\subsection*{Executive Summary}

For the reader's convenience, here we summarize the main results of our work in nontechnical terms, the technical 
computation is presented in Sec.~\ref{sec:methods}.
We have developed a general-relativistic, perturbative framework to construct equilibrium 
configurations of self-gravitating bodies with small (but otherwise generic) axisymmetric deformations away 
from spherical symmetry. Our approach is based on a framework recently developed for 
vacuum spacetimes~\cite{Raposo:2018xkf} and extends the latter to the case of matter fields, in particular perfect 
fluids, which provide an accurate description of the interior of cold NSs. 
We consider a spherical-harmonic decomposition of the spacetime and of the stress-energy 
tensor, and solve for Einstein's field equations in the interior of the star 
perturbatively in the deformations. The numerical solution in the stellar interior is 
then matched to the analytical solution known in the exterior~\cite{Raposo:2018xkf}, using appropriate boundary 
conditions (cf. Sec.\ref{sec:thin-shell}).
Spherical NSs deformed by an intrinsic small angular momentum (constructed in the seminal papers by Hartle and 
Thorne~\cite{Hartle:1967he,HT}) are a particular case of our general framework, which can be straightforwardly 
extended to arbitrary intrinsic multipole moments, to any perturbative order, and to different matter 
content. The framework is also similar to the linear perturbations of NSs due to tidal effects~\cite{Hinderer:2007mb}, 
although in our case there is no external source term such as a tidal field. For this reason we refer to the 
deformations as \emph{intrinsic} multipole moments, as opposed to the spin-induced or tidally-induced ones.
In the following we shall focus on the most interesting case of \emph{quadrupolar} deformations, being $\bar Q=Q/M^3$ 
the dimensionless quadrupole moment of a star with mass $M$.

In the absence of source terms (e.g. tidal fields, angular momentum, 
magnetic fields, shears, etc), the metric that describes a self-gravitating, perfect fluid with a quadrupolar 
deformation is discontinuous across the surface of the object. This shows that a (single) perfect-fluid star does not 
support deviations away from spherical symmetry in the static case \cite{1994CMaPh.162..123L}. 
However, here we show how an \emph{anisotropic} crust can support relatively large quadrupolar deformations.
The standard equation of state~(EoS) of a NS includes also the relatively low-density region of the 
crust~\cite{Lattimer:2004pg,Chamel:2008ca}, but typically assuming a perfect fluid for the latter, and therefore 
neglecting possible anisotropies. 
In reality the crust of a NS is crystallized, i.e. atomic nuclei form a lattice~\cite{NS,Chamel:2008ca,Caplan:2016uvu} 
whose elasticity can support asymmetric distributions~\cite{Baiko:2018jax}.
The (micro)physical processes of the crust are complex and model dependent. 
In an attempt to build a more general and less model-dependent configuration, we model the NS crust with a thin 
shell made of an anisotropic fluid and then match its properties with those of a more realistic crust model.
In the deformed case the discontinuity of the metric across the stellar radius, together with the junction conditions at 
the surface~\cite{Israel:1966rt,Visser:1995cc}, dictate the properties of the thin shell.

We have thus constructed explicitly 
a two-parameter family of NS equilibrium configurations characterized --~for each 
value of the central density~$\rho_c$~-- by an intrinsic quadrupole moment~$Q$, which is proportional to the anisotropy 
of the thin-shell crust. It is straightforward to add angular momentum $J$ perturbatively in our framework, so this 
family of solutions can be extended to a three-parameter model. For each value of $(\rho_c,J,Q)$, we can compute the 
other properties of the star, including the mass $M$ and radius~$R$. We stress that these solutions can be asymmetric 
even when nonrotating; for this reason we refer to them as \emph{deformed NSs}. An illustrative diagram of these 
solutions is presented in Fig.~\ref{fig:embedding}.

The junction conditions imply that the crust is made of an anisotropic fluid, whose surface energy density can be written in general as
\begin{equation}
\sigma(\theta)=\sigma_0+ \sigma_2{\bar Q} P_2(\cos\theta)\,, \label{sigmatheta}
\end{equation}
where the free parameter $\sigma_0\sim M_{\rm crust}/(4\pi R^2)$ is the surface density of the crust in the spherical 
configuration, $\sigma_2$ is the (normalized) amplitude of the axisymmetric perturbation to the surface density, and 
$P_\ell(\cos\theta)$ is the Legendre polynomial ($\ell=0,1,2,...$).
The anisotropy of the fluid on the thin shell can be measured by the difference
\begin{equation}
\Delta \gamma \equiv \gamma_\theta - \gamma_\phi = \bar Q \frac{3 \sin^2\theta}{8\pi R^2 
\sqrt{1-2M/R}}\left[\left[\xi\right]\right]\,, \label{deltagamma}
\end{equation}
%
between the surface pressure $\gamma$ along the two angular directions, where $\left[\left[\xi\right]\right]$ is the 
jump of the fluid displacement at the radius (see Sec.~\ref{sec:thin-shell} for details). 
Note that both the monopolar and quadrupolar components of the pressure 
are anisotropic.

In Fig.~\ref{fig:sigma2} we show the surface energy density $\bar{Q}\sigma_2$ (see Eq.~\eqref{sigmatheta}) and 
the surface pressure anisotropy $\Delta \gamma$ (see Eq.~\eqref{deltagamma}) at the equator as 
a function of the NS compactness for some relevant EoS. We normalize both quantities by the quadrupole value 
$|\bar{Q}|=0.1$, and by typical reference values, $\sigma_0\sim 10^{18}\,{\rm g\,cm}^{-2}$ and $\gamma_0\sim 
10^{17}\,{\rm g \, cm}^{-2}$, respectively, roughly corresponding to a crust containing $1\%$ of the total NS 
mass~\cite{Chamel:2008ca}. 
In particular, the values of the anisotropic pressure shown in Fig.~\ref{fig:sigma2} are compatible with the maximum 
allowed by elasticity of the crust, $\Delta \gamma/\gamma_0 \sim (0.005-0.04)Z^{4/3}$, where $Z\gtrsim 26$ is the atomic 
number of the ions in the crust (mostly iron in the outer crust~\cite{Chamel:2008ca}) and the prefactor depends on the 
type and direction of elastic deformations of the lattice~\cite{Baiko:2018jax}.
The main result that can be inferred from Fig.~\ref{fig:sigma2} is the fact that realistic values of the crust 
properties can sustain an intrinsic quadrupole as high as $\bar Q={\cal O}(0.1)$, with more conservative values being 
$\bar Q={\cal O}(0.01)$. We shall discuss the phenomenological implications of this intrinsic quadrupole in 
Sec.~\ref{sec:phenom}.

The rest of this paper is organized as follows. In the Sec.~\ref{sec:methods} we present the details of the 
technique used to construct deformed NSs, discussing the background configuration in Sec.~\ref{sec:equilibrium}. A 
discussion on the energy conditions is presented in Sec.~\ref{sec:energy_conditions}, whereas stability of the 
equilibrium configurations is discussed in Sec.~\ref{sec:linear_stability}. Finally, phenomenological effects of 
intrinsic deformations on the GW phenomenology are presented in Sec.~\ref{sec:phenom} and discussed on 
Sec.~\ref{sec:discussion}.

\section{Axisymmetric deformations of NSs: thin-shell crust}\label{sec:methods}

In this section we present the technical computation to construct NSs with perturbative deformations away from 
spherical symmetry.

\subsection{Equilibrium configurations}\label{sec:equilibrium}

\subsubsection{Formalism}

Our technique is an extension of the perturbative, general-relativistic framework recently 
developed to study generic axisymmetric departures from spherical symmetry in 
vacuum spacetimes~\cite{Raposo:2018xkf}.
We first provide the general framework for an arbitrary number of independent multipolar 
deformations, and then specialize it to the case in which the only 
nonvanishing deformations are due to the mass quadrupole moment (and, possibly, to the spin).

We consider a deformed metric of the form
\begin{equation}\label{eq:pertmetric}
g_{\mu\nu}=g_{\mu\nu}^{(0)}+\sum_{n=1}^\infty \epsilon^n h^{(n)}_{\mu\nu}\,,
\end{equation}
where $g_{\mu\nu}^{(0)}={\rm diag}\left\lbrace-e^{\nu(r)},1/(1-2m(r)/r),r^2,r^2\sin^2\theta\right\rbrace$ is the 
background metric which describes the 
spherically-symmetric star and can be obtained by solving for the 
Tolman-Oppenheimer-Volkoff 
equations~\cite{ShapiroTeukolsky}, $\epsilon$ is a small book-keeping parameter, and 
$h^{(n)}_{\mu\nu}$ is the 
deformation entering at order ${\cal O}(\epsilon^n)$. When only the mass quadrupole 
$Q$ (and, possibly, the angular moment ${J}$) are present at the leading order, the physical expansion 
parameters are the dimensionless quantities $\epsilon Q/{M}^3$ and $\epsilon {J}/{M}^2\equiv \epsilon\chi$, which are 
independent from one another.

Stationary and axisymmetric deformations can be expanded in a complete 
basis of Legendre polynomials:
%
 \begin{widetext}
\begin{align}
\label{eq:hpert}
&h_{\mu\nu}^{(n)}=
\sum_\ell\left(\begin{array}{cccc}
g_{00}^{(0)} H_0^{n\ell}P_\ell & 0 & 0 & h_0^{n\ell}P'_\ell \\
 0 & g_{rr}^{(0)} H_2^{n\ell}P_{\ell} & 0 &0\\
 0 &0 & r^2 K^{n\ell}P_{\ell}& 0\\ 
h_0^{n\ell }P'_\ell&0 &0 & r^2\sin^2{\theta} K^{n\ell}P_\ell
\end{array}\right)\,,
\end{align}
 \end{widetext} 
%
with $P_\ell=P_\ell(\cos\theta)$ and ${P_\ell}'=\frac{dP_\ell(\cos\theta)}{d\cos\theta}$. 
The parameter $\ell$ is related to the multipole moment sourced at each given order $n$ 
of the perturbative scheme. We separate the perturbations in two sets, 
according to how they transform under parity. The odd (or axial) sector 
contains only the radial function $h_0^{n\ell }(r)$, which is associated to the
\emph{current} multipole moments, $S_\ell$. The even (or polar) 
sector contains the radial functions $H_0^{n\ell}(r)$, $H_2^{n\ell}(r)$, and 
$K^{n\ell}(r)$, which are associated to the \emph{mass} multipole moments, $M_\ell$.
Note that we defined the mass quadrupole moment as $Q$; a more 
standard and general nomenclature is $Q\equiv M_2$~\cite{Geroch:1970cd,Hansen:1974zz}.

We consider Einstein's equations coupled to a perfect fluid that describes the interior 
of the star. The fluid's stress-energy tensor reads
\begin{equation}
 T^{\mu\nu}=(P+\rho)u^\mu u^\nu+P g^{\mu\nu}\,,\label{Tmunu}
\end{equation}
where 
\begin{eqnarray}
  P   &=& P^{(0)}(r) +\sum_{n=1}^\infty\sum_\ell \epsilon^n\delta P^{n\ell}(r) 
P_\ell\cos(\theta)\,,\\
 \rho&=&\rho^{(0)}(r)+\sum_{n=1}^\infty\sum_\ell \epsilon^n\delta \rho^{n\ell}(r) 
P_\ell\cos(\theta)\,,
\end{eqnarray}
are the pressure and the energy density of the fluid, respectively, the background values of which are denoted by
$P^{(0)}(r)$ and $\rho^{(0)}(r)$. The functions $\delta P^{n\ell}(r)$ and $\delta \rho^{n\ell}(r) $ are polar 
quantities. 
The four-velocity of a fluid element reads
\begin{equation}
 u^\mu = \frac{1}{\sqrt{-g_{tt}-2\epsilon\,\Omega 
g_{t\phi}-g_{\phi\phi}\epsilon^2\Omega^2}}\left\{1,0,0,\epsilon\,\Omega\right\}\,,
\end{equation}
where $\Omega$ is the fluid angular velocity and the normalization constant ensures that
$u^2=-1$.

By inserting the above decomposition for the metric and the fluid variables into 
Einstein's equations, $G_{\mu\nu}=8\pi T_{\mu\nu}$, and separating the angular dependence 
by using the orthogonality of the 
Legendre polynomials~\cite{Maselli:2015tta,Raposo:2018xkf}, we obtain a set of ordinary 
differential equations for the deformation 
functions $h^{(n)}_{\mu\nu}(r)$, $\delta P^{(n)}(r)$, and $\delta\rho^{(n)}(r)$. The 
system is closed by assuming a barotropic EoS, $P=P(\rho)$. 

In the exterior, the solution to the system can be found analytically at any given 
order and it depends on an arbitrary number of 
(mass and current) multipole moments~\cite{Raposo:2018xkf}.
The explicit values of the multipole moments can be obtained by matching the external 
solution to the internal one at the radius $R$ of the star, defined by $P^{(0)}(r=R)=0$ at the leading order. 
The internal solution is obtained numerically using a Runge-Kutta $4$-th order 
scheme with adaptive meshes~\cite{webpage}.

The couplings between multipoles follow the standard 
addition rules for angular momenta in quantum mechanics, so that if two modes 
with $\ell_1$ and $\ell_2>\ell_1$ are present at a given order in $\epsilon$, to 
the next order they will source multipole moments with $\ell$ such that 
$\ell_2-\ell_1\leq \ell\leq\ell_2+\ell_1$, provided some terms are not forbidden by parity
and equatorial-symmetry selection rules~\cite{Raposo:2018xkf}.
In the specific case of deformations sourced by $J$ and $Q$, only the $\ell=1$ axial 
deformation and the $\ell=2$ polar deformation are respectively present to ${\cal O}(\epsilon)$. 
Therefore, all induced multipole moments of this family of solutions can be written as a 
combination of terms sourced by the spin and by the quadrupole; to the leading order:
\begin{align}
\label{eq:inducedmassJM2}
 \bar{M}_{\ell} &= \sum_{p=0}^{\ell/2}\alpha_{p} \chi^{\ell-2p} \bar 
Q^{p}\,, \qquad {\rm even}~\ell\geq4\,,\nonumber\\
  \bar{S}_\ell &= \sum_{p=0}^{(\ell-1)/2}\beta_{p} 
\chi^{\ell-2p}\bar Q^{p}\,, \quad {\rm odd}~\ell\geq3\,,\nonumber
\end{align}
whereas ${M}_\ell=0$ and ${S}_\ell=0$ for odd $\ell$ and even $\ell$, respectively. In 
the above equation, $\bar M_\ell=M_\ell/M^{\ell+1}$ and $\bar S_\ell=S_\ell/M^{\ell+1}$ 
are the normalized multipole moments of degree $\ell$. The prefactors $\alpha_{p}$ and 
$\beta_{p}$ depend on the central density and on the EoS and have to be computed numerically.

The solution can be computed to any given order using the set of rules 
discussed above. Nonetheless, to linear order in the perturbation framework the solution is very 
simple since it is not affected by the nonlinearities of the field equations.

\subsubsection{Explicit external solution to leading order}

In the exterior of the object the solution can be found analytically~\cite{Raposo:2018xkf}. To the leading order it 
reads
\begin{widetext}
\begin{align}
&g_{tt}=-\left(1-\frac{2 M}{r}\right)-\frac{5 Q \left(2 M \left(2 M^3+4 M^2 r-9 M r^2+3 r^3\right)+3 r^2 (r-2 M)^2 \log 
\left(1-\frac{2 M}{r}\right)\right)}{8 \left(M^5 r^2\right)}P_2\,,\\
&g_{rr}=\left(1-\frac{2M}{r}\right)^{-1}-\frac{5 Q \left(2 M \left(2 M^3+4 M^2 r-9 M r^2+3 r^3\right)+3 r^2 (r-2 M)^2 
\log \left(1-\frac{2 M}{r}\right)\right)}{8 M^5 (r-2 M)^2}P_2\,,\\
&g_{\theta\theta}=r^2-\frac{5 Q r  \left(-4 M^3+\left(3 r^3-6 M^2 r\right) \log \left(1-\frac{2 M}{r}\right)+6 M^2 r+6 
M 
r^2\right)}{8 M^5}P_2\,,\\
&g_{\phi\phi}=g_{\theta\theta}\sin ^2\theta\,,\\
&g_{t\phi}=-\frac{2 J \sin ^2\theta }{r}\,.
\end{align}
\end{widetext}
%

\subsubsection{Thin-shell crust and junction conditions} \label{sec:thin-shell}
We can solve numerically the equations for the deformation in the interior of the star and match them with the above
analytical solution in the exterior using appropriate junction 
conditions~\cite{Israel:1966rt,Visser:1995cc,Uchikata:2015yma,Uchikata:2016qku}. 

The first junction condition imposes continuity of the induced metric $\eta_{ab}$ across the thin shell,
\begin{equation}
[[\eta_{ab}]]=0\,,
\end{equation}
where $[[X]]:=X_{\rm out}-X_{\rm in}$ denotes a jump of a generic quantity $X$ across the radius of the thin shell. To 
zeroth order in the deformations the first junction condition implies that 
\begin{equation}
e^{\nu(R)}=1-2 M /R\,,
\end{equation}
which is the usual continuity of time-time component of the metric, and can be imposed with a rescaling of the time 
coordinate in the interior of the star. To linear order this condition relates the jump of the metric functions with 
the quadrupole of the NS,
\begin{align}
&[[H_0^{1\ell}]]=-\left[\left[\frac{f^{\prime}}{f}\xi_2\right]\right] \,, \\
&[[K^{1\ell}]]=-\frac{2}{R}[[\xi_2]]\,,
\end{align}
where $f_{\rm out}=1-2M/r$, $f_{\rm in}=e^{\nu}$, and $\xi:=\xi_2 P_2(\theta)$ is the displacement of the thin shell.

The second junction condition relates the jump of the extrinsic curvature $K_{ab}$ with the stress-energy tensor,
\begin{equation}
S_{ab}=\frac{1}{8\pi}\left([[K_{ab}]]-\eta_{ab}[[K]]\right)\,,
\end{equation}
 with $K=\eta_{ab}K^{ab}$. The surface energy density $\sigma$ of the thin shell can be obtained as the eigenvalue of 
the stress energy tensor, namely
\begin{equation}\label{eq:stresshell1}
S^a_b u^b=-\sigma u^a\,.
\end{equation} 
The surface energy density can be written in general as in Eq.~\eqref{sigmatheta},
where $\sigma_0$ is the surface density of the thin shell in the spherically-symmetric configuration, whereas 
$\sigma_2$ 
is the 
amplitude of the axisymmetric perturbation to the surface density. By solving Eq.~\eqref{eq:stresshell1} we find the 
expressions for the surface density components in terms of the jump of metric functions. The zeroth order spherical 
case 
is particularly simple,
\begin{equation}
\label{eq:sigma_0}
\sigma_0=-\frac{1}{4\pi R}\left[\left[\left(g_{rr}^{(0)}\right)^{-1/2}\right]\right]\,.
\end{equation}
Obviously, in the spherical-symmetric case the continuity of the mass function implies the absence of any shell 
of matter. Conversely, assuming the existence of a thin shell with mass $M_{\rm crust}=\delta M:= M-m(R)$ 
implies $[[g_{rr}^{(0)}]]\neq0$ and the presence of a surface energy density given by Eq.~\eqref{eq:sigma_0}. For 
small thin-shell masses Eq.~\eqref{eq:sigma_0} becomes
\begin{equation}
\label{eq_sigma_0_2}
\sigma_0=\frac{1}{4\pi R^2}\frac{\delta M}{\sqrt{1-2M/R}}+{\cal O}(\delta M^2)\,,
\end{equation}
which reduces to the Newtonian case $\sigma_0=\delta M/(4\pi R^2)$ when $M/R\ll1$.

By using the projection tensor $q_{ab}:=\eta_{ab}+u_a u_b$, we  can define the projected stress-energy tensor of the 
thin shell,
\begin{equation}
\label{eq:projectedstress}
\gamma_{ab}=S^{cd}q_{ac}q_{bd}\,.
\end{equation}

Again, the spherical case is particularly simple and it reduces to the stress-energy tensor of a perfect fluid,
\begin{equation}\label{eq:isotropicshell}
\gamma_{ab}=\gamma_0q_{ab}\,,
\end{equation}
where $\gamma_0$ is the surface pressure of the shell. Comparing Eq.~\eqref{eq:projectedstress} with 
Eq.~\eqref{eq:isotropicshell} yields
\begin{equation}
\label{eq:}
\gamma_0=\frac{1}{8\pi R}\left[\left[\frac{1+\frac{R}{2} \frac{f'}{f}}{\sqrt{g_{rr}^{(0)}}}\right]\right]\,,
\end{equation}
which for small thin-shell masses can be reduced to
\begin{equation}
\gamma_0\sim \frac{M}{8 \pi  R^3}\frac{\delta M}{(1-2M/R)^{3/2}}+{\cal O}(\delta M^2)\,.
\end{equation}

To allow for quadrupolar deformations we will consider an anisotropic fluid for the matter of the thin shell, whose 
stress-energy tensor reads

\begin{equation}\label{eq:anysotropicshell}
\gamma_{ab}=\gamma^\theta k_{a} k_{b}+\gamma^\phi \chi_{a}\chi_b\,,
\end{equation}
where $k_a$ and $\chi_a$ are two normal vectors tangent to the thin shell, orthogonal to the fluid velocity $u_a$ and 
among themselves,
i.e. $\chi_a k^a=k_a u^a=\chi_a u^a=0$, $k_a k^a=1=\chi_a \chi^a$.
The latter five conditions do not specify the two vectors completely. There is one remaining degree of freedom which is 
related to the orientation of the two vectors with respect to the coordinate axis. One can fix this freedom by choosing 
to align $k_a$ with the $\theta$-axis and $\chi_a$ with the $\phi$-axis. With this choice $\gamma^\theta$ and 
$\gamma^\phi$ are the components of the surface pressure measured along the $\theta$ and $\phi$ directions, 
respectively. To calculate the pressure components it is useful to decompose each of them as 
$\gamma^i=\gamma^i_0+\gamma_2^i P_2(\theta)$, and solve Eqs.~\eqref{eq:projectedstress} and \eqref{eq:anysotropicshell} 
for the four independent pressure functions.

For example, the quadrupolar component of the surface energy density reads
\begin{widetext}
\begin{equation}
\sigma_2=\frac{1}{8\pi R}\left(\left[\left[\sqrt{1-2M/R}\, H_2 
\right]\right]-\left[\left[\frac{4+6M/R}{R\sqrt{1-2M/R}}\xi_2\right]\right]-\left[\left[R\sqrt{1-2M/R}\,K'\right]\right]
\right)\,.
\end{equation}
\end{widetext}
A similar (albeit more involved) relation can be derived for the quantity $\Delta \gamma$ defined in 
Eq.~\eqref{deltagamma}. This anisotropy $\Delta \gamma$ is required to support the quadrupolar deformation. When the 
surface pressures along the two angular directions on the shell are identical, the boundary conditions 
described above imply that the quadrupole moment must vanish.

\subsubsection{Examples of deformed NS configurations}
To summarize, the metric and fluid variables obtained numerically in the interior of the object are then matched to the 
external solution through the above junction conditions. 
The numerical solutions form a $3$-parameter family, which depends on the central 
density~$\rho_c$, the angular momentum $J$ (aligned with the symmetry axis, see 
Fig.~\ref{fig:embedding}) and the mass quadrupole moment $Q\equiv M_2$.

\begin{figure}[ht]
\centering
\includegraphics[width=0.49\textwidth]{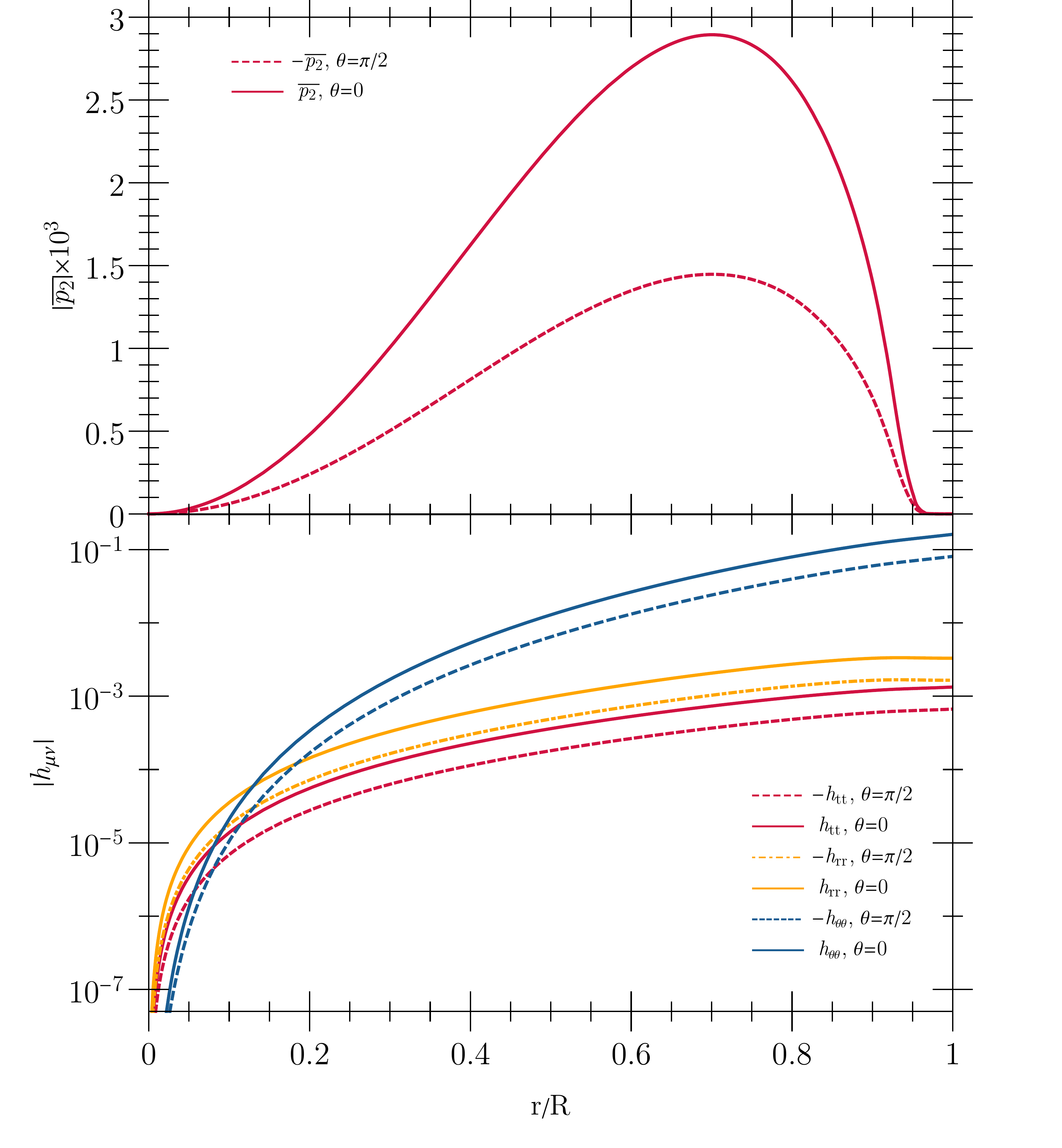}
\caption{{\bf Internal solutions.} Radial profiles of the fluid and metric variables in the interior of a deformed NS 
evaluated along the equatorial direction ($\theta=\pi/2$, dashed lines) and polar direction ($\theta=0$, solid 
lines). We show the profile 
of the pressure deformation (top panel) due to an intrinsic quadrupole deformation normalized by the central pressure 
of the object, $\bar{p}_2\equiv(P(r,\theta)-P_0(r))/P_0(0)$, and the metric deformation functions $h_{\mu\nu}\equiv 
g_{\mu\nu}-g_{\mu\nu}^{(0)}$ (bottom panel). The solution corresponds to a nonspinning prolate object with 
$M=1.4 M_\odot$ and $\bar{Q}=0.1$ to linear order in the deformation. To this order the deformations describing a 
nonrotating oblate 
object with the same magnitude of quadrupole ($\bar{Q}=-0.1$) are 
identical to those shown in this plot but with the opposite sign.}
\label{fig:interior}
\end{figure}

As a reference, in Table~\ref{tab:param} we 
provide some relevant quantities of the equilibrium configurations obtained with this procedure and assuming the AP4 
EoS. 
A representative example of the fluid and metric variables in the interior of a deformed NS is shown in 
Fig.~\ref{fig:interior}.

 \begin{table*}
 \begin{center}
 \resizebox{\textwidth}{!}{  
\begin{tabular}{ccc|ccccccc}
   \hline
   \hline
  $\rho_c~[10^{15}{\rm g}/{\rm cm}^3]$ & $\Omega/\Omega_K$ & $\bar{Q}/10^{-2}$ & $m(R)~[M_\odot]$ & 
$R~[{\rm 
km}]$ & $\bar I$ & ${\bar Q}\sigma_2/\sigma_0$ & $|\Delta \gamma/\gamma_0|^{\rm equator}$ & $\nu_\phi^{\rm ISCO}~[{\rm 
kHz}]$ & $\nu_\theta^{\rm 
ISCO}~[{\rm kHz}]$ \\
  \hline
  $1.540$	& 0	& 0	& $2.00$	& $11.0$	& $6.18$	& $0$ &$0$ & $1.10$ & 
$1.10$ \\
   $0.985$	& 0	& 0	& $1.40$	& $11.4$	& $11.1$	& $0$ & $0$ &$1.57$ & 
$1.57$ \\
   $0.875$	& 0	& 0	& $1.20$	& $11.5$	& $14.3$	& $0$ & $0$ &$1.84$ & 
$1.84$ \\
   \hline
    $1.540$	& $5\%$	& 0	& $2.00$	& $11.0$	& $6.18$	&$0$ & $0$ &
$1.13$ & $1.13$ \\
   $0.985$	& $5\%$	& 0	& $1.40$	& $11.4$	& $11.1$	& $0$ & $0$ &
$1.62$ & $1.61$ \\
   $0.875$	& $5\%$	& 0	& $1.20$	& $11.5$	& $14.3$	& $0$ & $0$ &
$1.90$ & $1.89$ \\
    $1.540$	& $10\%$& 0	& $2.00$	& $11.0$	& $6.18$	& $0$ & $0$ &
$1.17$ & $1.15$ \\
   $0.985$	& $10\%$& 0	& $1.40$	& $11.4$	& $11.1$	& $0$ & $0$ &
$1.67$ & $1.65$ \\
    $0.875$	& $10\%$& 0	& $1.20$	& $11.5$	& $14.3$	& $0$ & $0$ &
$1.96$ & $1.94$ \\
   \hline
   $1.540$	& 0	& $1$	& $2.00$	& $11.0$	& $6.18$	& $0.297 $& $0.445$ &$1.10$ 
& $1.10$ \\   
   $0.985$	& 0	& $1$	& $1.40$	& $11.4$	& $11.1$	& $0.133$ & $0.200$ &$1.57$ 
& $1.57$ \\
   $0.875$	& 0	& $1$	& $1.20$	& $11.5$	& $14.3$	& $0.0967$ & $0.145$ &$1.84$ 
& $1.84$ \\
   $1.540$	& 0	& $5$	& $2.00$	& $11.0$	& $6.18$	& $1.48 $& 
$2.22$ &$1.11$ & $1.11$ \\
   $0.985$	& 0	& $5 $	& $1.40$	& $11.4$	& $11.1$	& $0.667 $& $1.00$ &
$1.59$ & $1.58$ \\
   $0.875$	& 0	& $5$	& $1.20$	& $11.5$	& $14.3$	& $0.483$& $0.724$ &
$1.86$ & $1.85$ \\
   \hline
   \hline
 \end{tabular}
 }
 \end{center}
 \caption{{\bf Parameters of a deformed NS for AP4 EoS.} $\rho_c$ is 
the central energy 
density, $\Omega/\Omega_K$ is the angular velocity normalized by the mass-shedding limit $\Omega_K=\sqrt{M/R^3}$, 
$\bar{Q}$ is the dimensionless quadrupole moment, $M$ and $R$ are the stellar mass and radius, $\bar I$ is the 
normalized moment of inertia, ${\bar Q}\sigma_2/\sigma_0$ is the amplitude of the shell's surface density due to the 
quadrupole moment $\bar{Q}$ normalized by a typical surface density, $\sigma_0\approx 10^{18}\,{\rm g\,cm}^{-2}$, 
corresponding to $1\%$ of the stellar mass, $(\Delta \gamma/\gamma_{0})^{\rm equator}$ is the anisotropy of the shell 
at 
the equator normalized by the pressure of the spherical shell,
and $\nu_{\phi}^{\rm ISCO}$ and $\nu_{\theta}^{\rm ISCO}$ are the azimuthal frequency 
and the vertical epicyclic frequency at the innermost stable circular orbit (ISCO), see Sec.~\ref{sec:phenom}. 
}
\label{tab:param}
\end{table*}

\subsection{Energy conditions}\label{sec:energy_conditions}

Due to the broken spherical symmetry of the system, there are values of the parameter space for which the density might 
be negative, i.e, when $\bar{Q}\sigma_2 P_2(\theta)<-\sigma_0$ (see Eq.~\eqref{sigmatheta}). 
However, as shown in Fig.~\ref{fig:sigma2}, the surface density is always positive for a NS with $|\bar{Q}|\lesssim 0.1$ 
and compactness $M/R\lesssim 
0.15$. For the same range of quadrupole values, prolate stars have positive density for realistic ranges 
of compactness ($0.1\lesssim M/R\lesssim 0.2$), whereas more massive and compact ($M/R\gtrsim 0.16$) oblate stars 
require smaller quadrupoles to maintain positive density at the poles.

More generically, it is noteworthy to study all the energy conditions for the thin-shell fluid. The null energy 
condition reads $\sigma+\gamma_i\geq0$ (here $i=\theta,\phi$), the weak energy condition additionally requires 
$\sigma\geq0$; in addition to the latter condition, the strong energy condition also requires $\sigma+\sum_i 
\gamma_i\geq0$. Finally, the dominant energy condition requires only $\sigma>|\gamma_i|$.
A detailed numerical exploration of the parameter space shows that all energy conditions are satisfied for any 
realistic ranges of compactness whenever $\bar Q \lesssim M_{\rm crust}/M$. For the adopted reference value $M_{\rm 
crust}/M\sim 0.01$, the condition $|\bar{Q}|\lesssim 0.01$ ensures that all energy conditions are satisfied 
for any compactness, whereas if $|\bar{Q}|\lesssim 0.1$ all energy conditions are satisfied for NSs with compactness 
$M/R\lesssim0.15$.

\subsection{Linear stability analysis}\label{sec:linear_stability}

Performing a linear-stability analysis of deformed NSs is a challenging task, owing to the lack of symmetry of the 
background equilibrium solutions. To deal with this problem, we adopted a multi-parameter perturbative scheme, 
extending the techniques recently developed to study perturbations of slowly-rotating compact 
objects~\cite{Pani:2013pma,Pani:2015hfa,Pani:2015nua}. The computation can be summarized in the following 
steps~\cite{Pani:2013pma}: (i)~consider a complete set of small perturbations to the metric and the fluid variables, 
expanded in a basis of spherical harmonics $Y_{\ell m}(\theta,\phi)$; (ii)~solve for the dynamical equations 
perturbatively 
to leading order in the perturbation and to any given order in the background deformation; (iii)~Fourier-transform the 
perturbation functions and reduce the dynamical equations to a system of ordinary (radial) differential equations; 
(iv)~finally, 
solve for the system as an eigenvalue problem and obtain the characteristic modes of vibration. The eigenfrequencies 
$\omega$
are generically complex numbers and the sign of the imaginary part allows us to discriminate between stable and 
unstable configurations, owing to the $e^{-i\omega t}$ time dependence of the perturbations.

This standard procedure is made more involved by the broken symmetry of the background and by the presence of a thin 
shell. In an
axially symmetric background, perturbations with different values of the azimuthal
number $m$ are still decoupled, but the intrinsic quadrupole deformation introduces mixing between modes with opposite 
parity and different multipolar indices.

For heuristic purposes, let us first consider the dynamics of a test scalar field $\psi$ on the geometry of a deformed 
NS. By expanding $\psi=\sum_{\ell m}R_{\ell m}(r) r^{-1}Y_{\ell m}(\theta,\phi)e^{-i\omega t}$, the Klein-Gordon 
equation, 
$\square\psi=0$, reduces to
\begin{equation}
 \sum_{l=0}^\infty \left(A_\ell(r) Y_{\ell m} + D_\ell(r) P_2(\cos\theta) Y_{\ell m}(\theta,\phi)\right)=0\,, 
\label{system}
\end{equation}
where $A_\ell$ and $D_\ell$ are complicated functions of $R_{\ell m}(r)$ and of its derivatives up to second order, and 
they 
also depend on the frequency $\omega$ and on the multipolar index $\ell$. The function $A_\ell$ contains only 
quantities 
at 
zeroth order in the deformation, whereas the function $D_\ell$ is proportional to the quadrupolar deformation. Using 
the 
properties of the spherical harmonics~\cite{Pani:2013pma}, the above equation can be written as a system involving 
couplings between modes with harmonic index $\ell$ and $\ell\pm2$:
\begin{align}
 \frac{2}{3}A_\ell&+\left({\cal C}_{\ell +1}^2+{\cal C}_{\ell}^2-\frac{1}{3}\right) D_\ell\nn\\
 &+{\cal C}_{\ell}{\cal C}_{\ell-1}D_{\ell-2}+{\cal C}_{\ell +1}{\cal C}_{\ell+2}D_{\ell+2}=0\,, 
\label{systemcoupl}
\end{align}
where ${\cal C}_\ell\equiv \sqrt{\frac{\ell^2-m^2}{4\ell^2-1}}$. Formally, the above equation forms an infinite 
cascade of coupled 
ordinary differential equations since each $\ell$ mode is coupled to $\ell\pm2$. However, assuming only $\ell$-led 
perturbations 
to the leading order, the coupling to $\ell\pm2$ modes is subleading (because $D_{\ell\pm2}={\cal O}({\bar Q}^2)$) and 
Eq.~\eqref{systemcoupl} reduces to
\begin{equation}
 A_\ell+\frac{3}{2}\left({\cal C}_{\ell +1}^2+{\cal C}_{\ell}^2-\frac{1}{3}\right) D_\ell=0\,, \label{KG}
\end{equation}
which is now a single differential equation for the variable $R_{\ell m}$ (for each $\ell$). 

Remarkably, for radial perturbations ($\ell=m=0$) the second term in Eq.~\eqref{KG} vanishes identically and therefore 
radial perturbations are described by an equation of the form $A_0=0$, which coincides with the spherically-symmetric 
case. For $\ell>0$, this property is lost and the second term in Eq.~\eqref{KG} yields some corrections.
In this case the perturbation equation can be reduced to a Schr\"oedinger-like equation of the form $d^2 
R_{\ell m}/dx^2+(\omega^2-V_{\ell m})R_{\ell m}=0$, where $x$ is a new coordinates and the effective potential acquires 
a 
correction proportional to $Q$. In the exterior, the form of the potential can be computed analytically.
An analysis of the potential and corresponding eigenvalue problem shows that there are no unstable modes for any $\ell$.

Let us now turn to the more interesting case of gravitational perturbations. This is much more involved 
due to the tensorial nature of the perturbations and the coupling to the fluid modes. Nonetheless, in the radial case 
we 
can still obtain a system of equations which is formally equivalent to Eq.~\eqref{KG}, where now $A_\ell$ and $D_\ell$ 
are vector functions of all metric and fluid perturbations. Thus, we obtain the remarkable result that, to linear 
order in the quadrupolar deformation, radial perturbations of a deformed NS are governed by the same equations as in 
the spherical case. 
In particular, this implies that deformed NSs are stable under radial perturbations for configurations below the 
maximum 
mass~\cite{ShapiroTeukolsky}. 
The case of axial dipolar ($\ell=1$) perturbations and of axial and polar perturbations with $\ell\geq2$ is much more 
involved, due to the coupling among different 
sectors, and is left for future work. 

We note that, within our perturbative framework, the characteristic frequencies 
acquire small corrections proportional to $Q$. Therefore, only marginally stable modes in the spherical case can turn 
unstable. Instabilities, if they exist, can only come from the fluid axial sector for $\ell\geq2$, which contains a 
zero mode ($\omega=0$) in the spherical case. This mode can turn unstable due to the term proportional to $Q$, 
similarly to the case of the r-mode instability of slowly-spinning 
NSs~\cite{Andersson:1997xt,Friedman:1997uh,Andersson:2000mf}, 
for which the unstable mode has $\omega=m\Omega$. If present, an r-mode-like instability might remove part of the 
intrinsic quadrupole moment until the mode is saturated.

\section{Phenomenological implications}\label{sec:phenom}

As previously shown, a thin crust with reasonable values of anisotropy can sustain an intrinsic
(dimensionless) quadrupole moment as high as $\bar Q={\cal O}(0.1)$, with more conservative values being $\bar Q={\cal 
O}(0.01)$. 
The magnitude of the intrinsic quadrupole should be compared with the spin-induced quadrupole moment of a 
slowly-spinning NS, which scales quadratically with the dimensionless spin parameter, $\bar Q_{\rm spin} =-
\gamma \chi^2$, where $\gamma\approx 4\div7$ for compact stars 
with $M\approx 1.4M_\odot$, the precise number depending on the EoS~\cite{Hartle:1967he,HT,Yagi:2013bca,Yagi:2016bkt}. 
Since the spin $\chi$ of a NS is typically small~\cite{Dietrich:2015pxa} (roughly $\chi\approx0.1$ for the fastest 
millisecond pulsars and likely much smaller for old NSs in coalescing binaries detectable by 
LIGO/Virgo), the spin-induced quadrupole moment is at most $|\bar Q_{\rm spin}|\approx 7\times 10^{-2}$ and typically 
smaller. 
This suggests that any putative quadrupole moment of an old NS might predominantly natal rather than spin-induced.
If this is the case, a deformed NS might provide a better model for the external spacetime of the body.

Any quantity $X$ of a spinning, deformed NS contains independent $J$-induced 
and $Q$-induced corrections; schematically, up to second order in the 
deformation~\cite{Raposo:2018xkf}, 
\begin{equation}
 X = X_0 + X_{10} \chi  +  X_{01} \bar Q+  X_{20} \chi^2+  X_{02} \bar Q^2+ X_{11} 
\chi \bar Q\,, \label{schematic}
\end{equation}
where $X_0$ is the value of the corresponding spherically-symmetric star and $X_{ij}$ are corrections that only 
depend on the EoS and on the central density of the star. When $Q=0$, 
we recover the well-known case of a slowly-spinning NS~\cite{Hartle:1967he,HT}.

\subsection{Corrections of geodesic frequencies due to an intrisic quadrupole}

The external metric can be used to study how the spacetime of a deformed NS is 
affected by its intrinsic quadrupole. All geodesic quantities 
--~including the ISCO and the epicyclic 
frequencies~-- acquire corrections as in Eq.~\eqref{schematic}, which affect 
properties such as the innermost location of an accretion disk and, in turn, 
the corresponding electromagnetic flux from accreting low-mass X-ray 
binaries, whose signal originates very deep in the
gravitational field of the accreting object, at distances down to a few
gravitational radii~\cite{vanderKlis:2000ca}.
To linear order, the azimuthal 
frequency~($\nu_\phi$) and the vertical epicyclic frequency~($\nu_\theta$) at the ISCO read (see 
Appendix~\ref{app:epicyclic})
\begin{eqnarray}
 \nu_\phi^{\rm ISCO} &\approx& 1.57 \left(1+0.75\chi+0.23\bar Q\right) \left(\frac{1.4 
M_\odot}{M}\right)\,{\rm kHz}\,, \\
 \nu_\theta^{\rm ISCO} &\approx& 1.57 \left(1+0.61\chi+0.17\bar Q\right) \left(\frac{1.4 
M_\odot}{M}\right)\,{\rm kHz} \,,
\end{eqnarray}
independently of the NS EoS.
When $\bar Q\approx 0.1$, these frequencies can differ by a few percent relative to the 
spherical case, leading to deviations in the emitted flux of the same order. The 
quadrupolar correction is larger than the spin-induced linear term whenever $\bar 
Q\gtrsim 0.18\left(\frac{\chi}{0.05}\right)$.

\begin{figure}[th]
\centering
\includegraphics[width=0.475\textwidth]{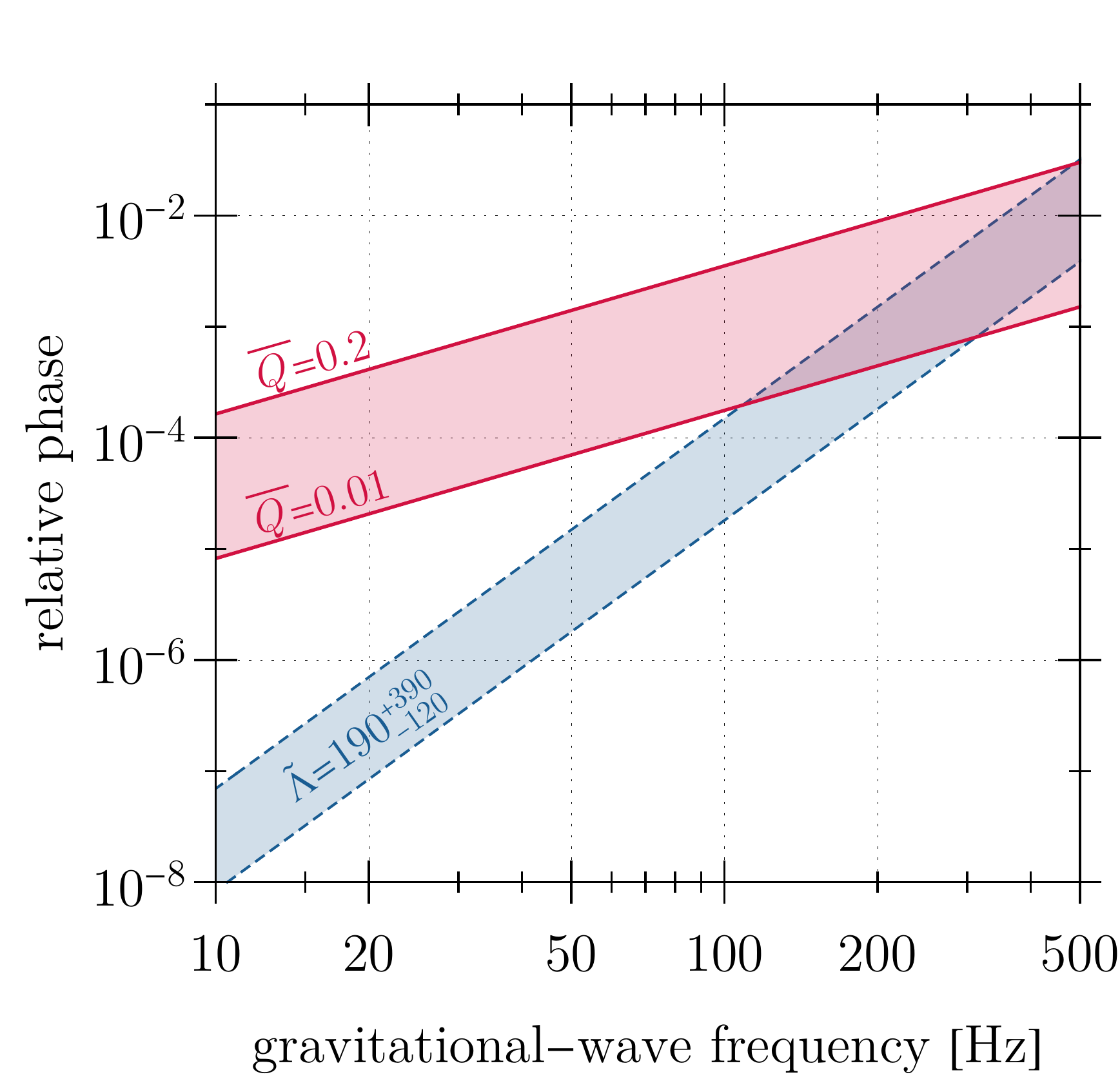}
\caption{{\bf GW phase.} Phase 
contribution, $\phi_{\rm quadrupole}$, for the coalescence of two NSs with intrinsic quadrupole moment relative 
to the leading-order Newtonian term in the LIGO band. 
The red band corresponds to the range $\bar Q\in (0.01,0.2)$.
The effect of the intrinsic deformation is larger than that 
of the tidal deformability (blue band) below $100\div 200\,{\rm Hz}$ even for (normalized) intrinsic quadrupole moment 
as small $\bar Q\approx0.01$, and it always dominates at lower frequencies 
or for higher deformations. We use reference values $m_1=m_2=1.4M_\odot$, and tidal deformability $\tilde 
\Lambda=190_{-120}^{+390}$~\cite{Abbott:2018exr,De:2018uhw}.}
\label{fig:phase}
\end{figure}

\subsection{GW phase corrections due to an intrisic quadrupole}

NSs in compact binaries are assumed to be spherically symmetric at 
large orbital distance $d$, whereas they are deformed during the 
coalescence due to tidal interactions~\cite{Flanagan:2007ix}. The tidally-induced quadrupole 
moment is proportional to the tidal field $\sim M/d^3$; the leading-order tidal 
correction to the GW phase enters at the fifth post-Newtonian\footnote{The post-Newtonian approach is a 
weak-field/slow-velocity expansion 
of Einstein's equations, where the expansion parameter is the orbital velocity 
$v\ll1$. A $n$-th post-Newtonian correction to the 
GW phase corresponds to a term that is suppressed by $v^{2n}$ relative to 
the leading-order, $\sim v^{-5}$, contribution.} 
order~\cite{Blanchet:2006zz}, i.e. $\phi_{\rm tidal}\sim v^5$, and it is proportional 
to the tidal deformability $\tilde\Lambda$, which characterizes the size of the tidally 
induced quadrupole deformations of the two stars~\cite{Flanagan:2007ix,GW170817}.

On the other hand, an intrinsic quadrupole moment of the two inspiraling bodies affects 
the GW phase already at second post-Newtonian order~\cite{Blanchet:2006zz,Krishnendu:2017shb}. Adapting the results for 
a spin-induced quadrupole moment, we find that a generic quadrupole moment gives the following GW phase contribution
\begin{equation}
 \phi_{\rm quadrupole} = \frac{75}{64} \frac{\left(m_1^2 \bar Q_1+m_2^2 \bar 
Q_2\right)}{ 
m_1 m_2}\frac{1}{v}\simeq \frac{75}{32} \frac{\bar Q}{v}\,,\label{phase}
\end{equation}
where in the last step we assumed the same masses~($m_1=m_2=M$) and the same intrinsic 
quadrupole moment~($\bar Q_1=\bar Q_2= \bar Q$) for the two stars.
As long as $\bar Q\approx 0.01$ or larger, the effect of the intrinsic 
quadrupole will dominate over the tidal term, especially at low 
frequencies~(Fig.~\ref{fig:phase}). To the best of our knowledge, 
this effect has never been considered before, but can dramatically affect the parameter 
estimation of NS binaries and significantly modify the constraints on the NS 
EoS~\cite{GW170817,Abbott:2018exr,De:2018uhw}.

\section{Discussion}\label{sec:discussion}

Although our analysis is based on some simplified assuptions, it suggests that intrinsic deformations in NSs could have 
important implications for high-energy astrophysics, GW astronomy, and nuclear physics. Deformed NSs 
might provide a more accurate description for the interior of NSs in binary systems and, in 
turn, for all the strong-gravity phenomena which are analyzed to infer the NS EoS.

It is difficult to estimate the degree of deformation expected for NSs formed in 
realistic situations, e.g.
in a merger, since the latter is often simulated with simplified initial data, with 
some assumed degree 
of symmetry, and the evolution is typically stopped before the remnant has reached a 
truly stationary configuration. More importantly, numerical simulations typically use perfect fluids, which cannot 
sustain deformations away from spherical symmetry in the static case~\cite{1994CMaPh.162..123L}. 
Although the (micro)physics of a NS interior is very complex~\cite{NS,Chamel:2008ca,Caplan:2016uvu}, deformations might 
be the rule rather than the exception~\cite{Baiko:2018jax}. Indeed, relatively large quadrupole moments can be sustained 
even by modest anisotropic stresses, as we showed. Another example is the case of self-gravitating elastic compact 
objects. In general, even in the static case, these objects do not need to be spherical since deformations from 
spherical symmetry can be supported by shears in the material \cite{Andersson:2006ze,Andersson:2008yf}. A very 
interesting extension of our work is to construct deformed NSs sustained by elasticity beyond the thin-shell formalism 
adopted here.

We have explicitly focused on the case of an anisotropic crust, but nonspherical deformations might also be sourced by 
other processes, for example by a magnetic field.
Relativistic models of axisymmetric, highly-magnetized NSs have been studied in great
detail, see e.g.~\cite{Ioka:2003nh,Colaiuda:2007br,2012MNRAS.427.3406F,2014MNRAS.439.3541P,2015MNRAS.447.3278B,
1995A&A...301..757B} . The quadrupole moment induced by the magnetic field $B$ can be parametrized as
 \begin{equation}
 \bar{Q} = -\beta \frac{\pi B^2 R^{8}}{\mu_0 I M^3}\approx -0.01 \beta \frac{B_{16}^2 R_{12}^8}{I_{45} M_{1.4}^3}\,, 
\label{Qbarmagnetic}
\end{equation}
where $B_{16}=\frac{B}{10^{16}\,{\rm Gauss}}$, $R_{12}=\frac{R}{12\,{\rm km}}$, $M_{1.4}=\frac{M}{1.4 M_\odot}$, 
$I_{45}=\frac{I}{10^{45}\,{\rm g \, cm}^2}$, $I$ 
is the moment of inertia, $\mu_0$ is the vacuum permeability, and 
$\beta\sim {\cal O}(1)$ is the magnetic distortion factor which measures to what extent a star can be deformed by the 
magnetic field. Note the very strong dependence on the radius of the star $R$ and the quadratic dependence on the 
magnetic field $B$. By computing $M$, $R$, and $I$ for a family of NSs with some tabulated EoS, one can check that the 
magnetic-induced $\bar{Q}$ is of the order of what given in Eq.~\eqref{Qbarmagnetic} for all configurations, and 
therefore very small unless the magnetic field is extreme.
In the case of deformation sourced by a magnetic field, the exterior of the star is not vacuum, so strictly speaking 
our solution in the exterior is not valid. 
However, the magnetic field is mainly dipolar and decays as $r^{-3}$. Sufficiently far from the star, the vacuum 
solution is a good approximation and the metric can be matched to the analytical one written in terms of the multipole 
moments~\cite{Raposo:2018xkf}.

Observational bounds on parametrized 
corrections to the post-Newtonian coefficient at second order from binary-NS coalescence
GW170817~($\delta \varphi_2\lesssim3.5$ at $90\%$ confidence level~\cite{Abbott:2018lct}) 
can be directly translated --~using Eq.~\ref{phase}~-- into a bound on the intrinsic 
quadrupole moments of the binary components, yielding $\bar 
Q\lesssim 0.14$ (assuming equal masses). This justifies our perturbative treatment but 
does not exclude that the binary components of GW170817 had some significant 
intrinsic deformation. 
This also suggests that a putative measurement $\delta\varphi_2\neq0$ 
in the future might be due to an intrinsic deformation of the NSs rather than to a 
fundamental departure from general relativity.

An outstanding issue concerns the stability of deformed NSs at equilibrium. Performing a linear stability analysis is 
particularly challenging due to the broken symmetry of the equilibrium configuration. Nevertheless, such an analysis 
can 
be performed using advanced perturbation-theory techniques developed for slowly-spinning compact 
objects~\cite{Pani:2013pma,Pani:2015hfa,Pani:2015nua}. 
A preliminary analysis shows that deformed NSs are stable against radial 
perturbations for masses below the maximum mass, exactly as spherically-symmetric NSs~\cite{ShapiroTeukolsky} and other 
thin-shell objects like gravastars~\cite{Uchikata:2016qku}.
Within our perturbative framework, possible instabilities might only come from (axial) nonaxisymmetric 
fluid perturbations, which are the only ones containing a \emph{zero mode} in the spherical case. This mode can turn 
unstable due to 
the correction proportional to $Q$, similarly to the case of the r-mode instability of slowly-spinning 
NSs~\cite{Andersson:1997xt,Friedman:1997uh,Andersson:2000mf}. If present, an r-mode-like instability might remove part 
of the intrinsic quadrupole moment until the mode is saturated by GW emission or other mechanisms.

To second order in the deformations, the coupling between $J$ and $Q$ moments induces 
novel multipole moments through the standard angular-momentum addition 
rules~\cite{Raposo:2018xkf}. In particular, it induces a mass hexadecapole and a current 
octupole, as well as a shift of the stellar mass and radius, angular momentum, and mass 
quadrupole moment. These corrections are quadratic in the deformation and therefore 
subleading. For example, for $\bar Q\approx 0.1$ in the nonspinning case the NS mass acquires a correction 
approximately at the percent level, which is comparable to the mass of the crust~\cite{Chamel:2008ca}.
On the other hand, higher-order corrections might be important to model highly-deformed 
stars more accurately.

To summarize, using a simple thin-shell model, we showed that anisotropic crust stresses can 
support significant quadrupolar deformations while satisfying all energy conditions. Based on this model, we argue that 
if NSs display some departure from spherical 
symmetry of the order of at least a few tenths of a percent, their intrinsic quadrupole moment would 
introduce corrections that are more important than the spin and tidal deformations. We 
advocate the urgency of quantifying NS intrinsic deformations and of including them in the 
parameter estimation of electromagnetic and GW signals from isolated and 
binary NSs.

\begin{acknowledgments}
We are indebted with Carlos Palenzuela for initial collaboration that lead to this work.
We are also grateful to Vitor Cardoso, Valeria Ferrari, Bruno Giacomazzo, Kostas Kokkotas, Massimo Mannarelli, and 
Luigi Stella for interesting discussion.
P.P. acknowledges financial support provided under the European Union's H2020 ERC, Starting 
Grant agreement no.~DarkGRA--757480, and under the MIUR PRIN and FARE programmes (GW-NEXT, CUP:~B84I20000100001).
The authors would like to acknowledge networking support by the COST Action CA16104 and 
support from the Amaldi Research Center funded by the MIUR program ``Dipartimento di 
Eccellenza'' (CUP: B81I18001170001).
\end{acknowledgments}

\begin{widetext}
\appendix

\section{Epicyclic frequencies} \label{app:epicyclic}

The epicyclic frequencies can be computed for a generic stationary, axisymmetric 
spacetime~\cite{Maselli:2014fca}. Setting the spin to zero, to the leading order in $\bar 
Q$, the azimuthal frequency and the vertical and radial epicyclic frequencies read
\begin{eqnarray}
 \nu_\phi     &=&\frac{1}{2\pi M x^{3/2}}\left(1+ \Delta_\phi \bar Q\right) \,, \\
 \nu_\theta   &=&\frac{1}{2\pi M x^{3/2}}\left(1+ \Delta_\theta \bar Q\right)\,, \\
 \nu_r        &=&\frac{\sqrt{x-6}}{2\pi M x^{2}}\left(1+ \Delta_r\bar Q\right) \,, 
\end{eqnarray}
respectively, where $x\equiv r/M$ and
\begin{eqnarray}
 \Delta_\phi   &=&\frac{-15 (x-2) x \left(x^3-2\right) \log \left(\frac{x-2}{x}\right)+10 
x (x (2-3 (x-1) x)+8)-60}{32 (x-2) x} \,, \nonumber\\
 \Delta_\theta &=&\frac{5}{32} \left(6 x (2 x-7)-\frac{12}{(x-2) x}+3 (2 x-1) (x-2)^2 
\log 
\left(\frac{x-2}{x}\right)+34\right)\,,\nonumber \\
 \Delta_r     &=& \frac{10 (48+x (30+x (26+x (3 (25-4 x) x-127))))-15 (x-2)^2 x^2 (x 
(4 x-13)-2) \log\left(\frac{x-2}{x}\right)}{32(x-6) (x-2) x}\,. \nonumber
\end{eqnarray}

For completeness, we also provide the leading-order quadrupolar corrections to the ISCO 
radius in closed form, including also the well-known spin term at the linear order:
\begin{equation}
 r_{\rm ISCO} = 6M\left[1-\frac{2\sqrt{2}}{3\sqrt{3}}\chi+\left(\frac{9325}{96}-480 \coth 
^{-1}(5)\right)\bar Q\right]\,,
\end{equation}
and the corresponding analytical expressions for $\nu_\phi$ and 
$\nu_\theta$ at the ISCO:
\begin{eqnarray}
\label{eq:freq_phi}
 \nu_\phi^{\rm ISCO} &=& \frac{1}{12 \pi\sqrt{6} M}\left[1+\frac{11}{6 
\sqrt{6}}\chi+\frac{5}{32} \left(5892 \coth ^{-1}(5) - 1193 \right)\bar Q\right]\,,\\
\label{eq:freq_theta}
 \nu_\theta^{\rm ISCO} &=&\frac{1}{12 \pi\sqrt{6} M}\left[1+\frac{\sqrt{3}}{2 
\sqrt{2}}\chi+\left(555 \coth ^{-1}(5)-\frac{3595}{32}\right)\bar 
Q\right] \,.
\end{eqnarray}
An approximate version of the above equations has been presented in the main text.
\end{widetext}

\bibliography{refs}
\end{document}